%% file: Wireless_network_control_of_interacting_Rydberg_atoms__arXiv_edition.tex
\documentclass[aps,prl,twocolumn,superscriptaddress,groupedaddress,showpacs]{revtex4}

\usepackage[english]{babel}
\usepackage{dsfont} 

\usepackage{amsmath}
\usepackage{amsthm}
\usepackage{amssymb}
\usepackage{booktabs}
\usepackage{hyperref}
\hypersetup{
  colorlinks   = true,  
  urlcolor     = blue,  
  linkcolor    = blue,  
  citecolor    = red    
}
\usepackage{natbib}
\usepackage{pgfplots}
\usepackage{siunitx}
\usepackage{subfigure}
\usepackage{tikz}
\usetikzlibrary{arrows,automata,backgrounds,decorations,positioning,shapes,snakes}
\usepackage{url}
\usepackage{xifthen}

\newcommand{\iterand}[2]{ #1^{[#2]} }
\newcommand{\itr}[2]{ \iterand{#1}{#2} }
\newcommand{\vect}[1]{ \boldsymbol{#1} }
\newcommand{\vectones}[1]{ \vect{1}_{#1} }

\newcommand{\elementaryVectorNoSize}[1]{ \vect{e}_{#1} }
\newcommand{\vectComponent}[2]{ #1_{#2} }
\newcommand{\vC}[2]{ \vectComponent{#1}{#2} }
\newcommand{\vectInLine}[1]{ ( #1 )^{\mathrm{T}} }
\newcommand{\matrixElement}[3]{ {#1}_{#2,#3} }

\newcommand{\transpose}[1]{ #1{}^{\mathrm{T}} }

\newcommand{\criticalpoint}[1]{  #1^{\textnormal{opt}} }

\newcommand{\cardinality}[1]{ | #1 | }

\newcommand{\indicator}[1]{ \mathds{1} [ #1 ] }

\newcommand{\process}[2]{ \{ #1 \}_{ #2 } }

\newcommand{\naturalNumbersPlus}{ \mathbb{N}^+ }
\newcommand{\naturalNumbersZero}{ \mathbb{N}_0 }
\newcommand{\realNumbers}{ \mathbb{R} }

\newcommand{\sA}{ {\ensuremath{\sigma}} } 
\newcommand{\sB}{ {\ensuremath{\eta}} } 

\newcommand{\svA}{ {\vect{\sA}} }
\newcommand{\svB}{ {\vect{\sB}} }

\newcommand{\stateSpace}{ \mathcal{S} }
\newcommand{\itstep}[1]{\iterand{a}{#1}}

\newcommand{\normalizationConstant}{Z}

\newcommand{\ket}[1]{ | #1 \rangle }

\newcommand{\RabiFrequency}{ \Omega } 
\newcommand{\decayRate}{ \Gamma }

\newcommand{\refFigure}[1]{{\textrm{Fig.~\ref{#1}}}}

\newcommand{\refEquation}[1]{{\textrm{Eq.~(\ref{#1})}}}

\begin{document}

\title{Wireless network control of interacting Rydberg atoms}

\author{Jaron \surname{Sanders}}
\email{jaron.sanders@tue.nl}

\author{Rick \surname{van Bijnen}}
\altaffiliation[Present address: ]{Max Planck Institute for the Physics of Complex Systems, 01187 Dresden, Germany}

\author{Edgar \surname{Vredenbregt}}

\author{Servaas \surname{Kokkelmans}}
\affiliation{Eindhoven University of Technology, P.~O.~Box 513, 5600 MB Eindhoven, The Netherlands}

\date{April 29, 2014}


\begin{abstract}
We identify a relation between the dynamics of ultracold Rydberg gases in which atoms experience a strong dipole blockade and spontaneous emission, and a stochastic process that models certain wireless random-access networks. We then transfer insights and techniques initially developed for these wireless networks to the realm of Rydberg gases, and explain how the Rydberg gas can be driven into crystal formations using our understanding of wireless networks. Finally, we propose a method to determine Rabi frequencies (laser intensities) such that particles in the Rydberg gas are excited with specified target excitation probabilities, providing control over mixed-state populations.
\end{abstract}

\maketitle


Stochastic processes play a ubiquitous role in interacting particle systems. Glauber initiated a study of the stochastic Ising model in 1963 \cite{Glauber63}, and similar models are actively investigated in probability theory, often applied to very different systems \cite{Liggett10,liggett_interacting_1985}. The two seemingly disparate interacting particle systems we study in this Letter, are a gas of ultracold Rydberg atoms \cite{Gallagher} accompanied by a dissipative mechanism, and a wireless random-access network, made up of for example electronic transmitters in communication networks \cite{kelly_stochastic_1985}. It turns out that their dynamics can be described, under certain conditions, with the same equations. Indeed, Rydberg atoms exhibit a strong interaction, while simultaneously active transmitters would lead to interference at receivers, both resulting in complicated large-scale system behavior.

Rydberg gases consist of atoms that can be in either a ground state or an excited state with a high principal quantum number. When an atom is excited, the energy levels of neighboring atoms shift. This makes it unlikely for neighboring atoms to also excite, and we call this effect the dipole blockade \cite{Lukin01,Comparat10}. The dipole blockade is at the basis of quantum information and quantum gate protocols \cite{Jaksch00,Lukin01,Saffman10}, and also allows for a phase transition to ordered structures \cite{Weimer08}. Experimentally, the cNOT gate has been demonstrated \cite{Saffman10}, while also the first ordered Rydberg structures have been observed \cite{Schauss12}. Recent experiments are geared towards leveraging the dipole blockade to create Rydberg crystals, i.e.~formations of regularly spaced excited atoms. A proposed method is to use chirped laser pules \cite{Pohl10,Schachenmayer10,bijnen_adiabatic_2011}, and another utilizes a dissipation mechanism: specifically spontaneous emission \citep{honing_steady-state_2013}.

Nowadays, transmitters in wireless networks share a transmission medium through the use of distributed random-access protocols. We focus on wireless networks operating according to the CSMA protocol \cite{kleinrock_packet_1975}, which lets 
transmitters autonomously decide when to start a transmission based on the level of activity in their environment, usually estimated through measurements of interference and signal-to-noise ratios. If too many neighbors are sensed to be transmitting, the transmitter postpones its activation and tries again at a random later point in time. We see that transmitters experience blocking effects similar to the Rydberg dipole blockade, which sparked our original interest to compare their mathematical models \cite{sanders_interacting_2011}. Mathematical models of wireless networks were already being studied because of our increasing demands on our communication infrastructures, and we focussed our attention on stochastic models of CSMA that were originally considered in \cite{kelly_stochastic_1985,boorstyn_throughput_1987,kershenbaum_algorithm_1987}.

This Letter uses the fact that rate equations adequately describe the Rydberg gas when spontaneous emission is introduced to the model \cite{ates_many-body_2007}, and we interpret the rate equations as Kolmogorov forward equations \cite{grimmett_probability_2001} that describe the transient evolution of a stochastic model reminiscent of CSMA.


Regarding the Rydberg gas, we consider a gas of $N$ atoms in the $\mu$-Kelvin regime, to which we apply the frozen gas approximation by neglecting the kinetic energy of the system. The atoms are thus considered fixed at positions $\vectComponent{\vect{r}}{i} \in \realNumbers^3$ for $i = 1, \ldots, N$.
The ultracold atoms are subjected to two lasers with associated Rabi frequencies $\RabiFrequency_{\textrm{e}}, \RabiFrequency_{\textrm{r}}$, respectively, that facilitate excitation from the ground state $\ket{\textrm{g}}$ to an intermediate state $\ket{\textrm{e}}$, and from the intermediate state $\ket{\textrm{e}}$ to a Rydberg state $\ket{\textrm{r}}$. We also assume that the intermediate state decays with rate $\decayRate$, through spontaneous emission. In principle, detuning of the laser frequencies could be taken into account, but here we leave it out for simplicity.

\begin{figure}[!hbt]
\begin{center}
\subfigure{
\includegraphics[height=0.2575\columnwidth]{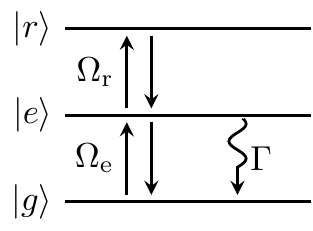}
}
\subfigure{
\includegraphics[height=0.2575\columnwidth]{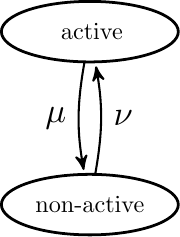}
}
\end{center}
\vspace{-0.75cm}
\caption{\textrm{(left) An atom can transition between a ground, intermediate and a Rydberg state. (right) A transmitter can change between nonactive and active.}}
\label{fig:Energy_diagram}
\vspace{-1em}
\end{figure}

The system description of a wireless random-access network is similar to that of the Rydberg gas, but with different terminology. A wireless random-access network can be modeled as consisting of $N$ transmitter-receiver pairs, and each transmitter can be either active ($1$) or nonactive ($0$). When active, a transmitter transmits data for an exponentially distributed time with mean $1 / \mu$. Similarly, a nonactive transmitter repeatedly attempts to become active after exponentially distributed times with mean $1 / \nu$. \refFigure{fig:Energy_diagram} summarizes our modeling assumptions thus far.

For a single atom, we can write down the optical Bloch equations. As in \cite{ates_many-body_2007}, we then conclude that if (i) the upper transition is much more weakly driven than the lower one ($\RabiFrequency_{\textrm{r}} \ll \RabiFrequency_{\textrm{e}}$), and (ii) the decay rate of the intermediate level is much larger than the Rabi frequency driving between $\ket{\textrm{e}}$ and $\ket{\textrm{r}}$ ($\RabiFrequency_{\textrm{r}} \ll \decayRate$), that then the excitation dynamics are described using the rate equation,
\begin{equation}
\frac{ d \vC{p}{1}(t) }{ dt } 
= \nu \vC{p}{0}(t) - \mu \vC{p}{1}(t). \label{eqn:Single_atom_rate_equation}
\end{equation}
Here, $\vC{p}{0}(t)$ and $\vC{p}{1}(t)$ denote the probabilities that the atom is (effectively) in the ground state or the Rydberg state, respectively. Furthermore,
\begin{equation}
\mu
= \frac{ 2 \decayRate \RabiFrequency_{\textrm{r}}^4  }{ ( \RabiFrequency_{\textrm{r}}^2 - 2 \RabiFrequency_{\textrm{e}}^2 )^2 + 2 \decayRate^2 ( \RabiFrequency_{\textrm{e}}^2 + \RabiFrequency_{\textrm{r}}^2 ) }, \textrm{ and }
\nu
= \frac{ \RabiFrequency_{\textrm{e}}^2 }{ \RabiFrequency_{\textrm{r}}^2 } \mu
\label{eqn:Rydberg_gas__Transition_rates_for_one_atom}
\end{equation}
denote the transition rates between the ground and Rydberg state. It is noteworthy that \refEquation{eqn:Single_atom_rate_equation} also describes the time evolution of a single, noninteracting transmitter. The $p_0(t)$ and $p_1(t)$ are then the probabilities that the transmitter is nonactive or active, respectively.

When dealing with many-particle systems, however, we have to take particle interactions into account. The atoms in Rydberg gases, and the transmitters in wireless networks, interact with each other. Specifically, if an atom is in the Rydberg state, other nearby atoms experience a dipole blockade \cite{bijnen_adiabatic_2011}. Transmitters that detect high levels of interference and low signal-to-noise ratios (because of their neighbors) postpone their activation.

We will model the dipole blockade, as well as the interference constraints on transmitters, using a unit-disk interference model. The unit-disk interference model involves the assumption that atoms (transmitters) within a distance $R$ of each other cannot simultaneously be in the Rydberg state (active). For Rydberg gases, this assumption is in line with measurements and simulations of pair correlation functions between atoms in the Rydberg state, which show a sharp cutoff when plotted as a function of the distance between the atoms \cite{Schauss12,Robicheaux05}. The collection of possible configurations 
is thus
\begin{equation}
\stateSpace 
= \bigl\{ \svA \in \{ 0, 1 \}^N \big| d( \vect{r}_i, \vect{r}_j ) > R \, \forall_{ i \neq j : \vC{\sA}{i} = \vC{\sA}{j}  = 1} \bigr\},
\end{equation}
and these configurations $\svA = \vectInLine{ \vectComponent{\sA}{1}, \ldots, \vectComponent{\sA}{N} }$ will be called feasible. The notation is such that if $\vC{\sA}{i} = 0$ or $1$, atom $i$ is in the ground or Rydberg state, respectively. Similarly, $\vC{\sA}{i} = 0$ or $1$ if transmitter $i$ is nonactive or active, respectively.  

\begin{figure}[!hbtp]
\begin{center}
\subfigure{
\includegraphics[height=0.309\columnwidth]{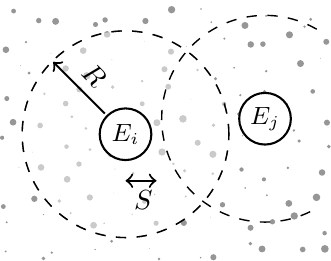}
}
\subfigure{
\includegraphics[height=0.309\columnwidth]{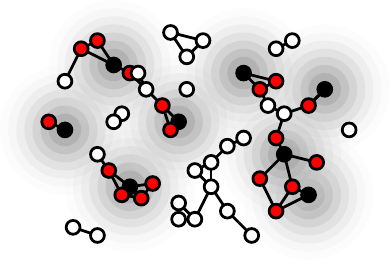}
}
\end{center}
\vspace{-0.75cm}
\caption{\textrm{The Rydberg blockade prevents atoms within a radius $R$ from becoming Rydberg atoms (left). This interaction can be described using an interference graph where edges indicate which neighboring particles would block each other, which is part of the wireless network model (right). Active transmitters (black) prevent neighboring transmitters (red) from becoming active. Non-neighboring nonactive transmitters (white) can become active.}}
\label{fig:Laser_regions_and_Rydberg_blockade}
\vspace{-1em}
\end{figure}

There are certainly practical differences between Rydberg gases and wireless networks. In wireless networks, every transmitter can have its own activation ($\vectComponent{\nu}{i}$) and deactivation rate ($\vectComponent{\mu}{i}$). To achieve the same effect in Rydberg gases, we will assume that the two-step laser can be split into $M \ll N$ spots with radius $S$, and that each spot $i = 1, \ldots, M$ has a different laser intensity $\vC{E}{i}$. Each laser spot contains a cluster of atoms, and with this setup, the atoms within each cluster may be subjected to a different Rabi frequency. We assume that $S \ll R$, so that we can treat each spot as being synonymous to one atom, and we will replace the symbol $M$ by $N$ for notational convenience. Each atom (spot) $i = 1, \ldots, N$ will thus experience its own transition rates $\vC{\nu}{i}$, $\vC{\mu}{i}$. \refFigure{fig:Laser_regions_and_Rydberg_blockade} summarizes the blockade effect, our assumptions on the laser spots, and the unit-disk interference model.

For both models, the probability of observing the system in state $\svA \in \stateSpace$ at time $t$, denoted by $\vC{p}{\svA}(t)$, is described by the master equation
\begin{equation}
\frac{ d \vC{p}{\svA}(t) }{ dt }
= \sum_{ \svB \in \stateSpace } \matrixElement{Q}{\svA}{\svB} \vC{p}{\svB}(t) \label{eqn:Rate_equations},
\end{equation}
where $Q$ denotes a transition rate matrix. The master equation \refEquation{eqn:Rate_equations} can be interpreted as a Kolmogorov forward equation, which characterizes a Markov process \cite{grimmett_probability_2001,kelly_reversibility_2011}. 

The off-diagonal elements of $Q \in \realNumbers^{ \cardinality{\stateSpace} \times \cardinality{\stateSpace} }$ describe the dynamics of this stochastic process. Denoting the $N$-dimensional vector with a one in the $i$th position by $\elementaryVectorNoSize{i}$, we have that when the system is in state $\svA \in \stateSpace$, it jumps to states $\svA + \elementaryVectorNoSize{i}$, $i = 1, \ldots, N$, with rate
$
\matrixElement{Q}{\svA}{\svA + \elementaryVectorNoSize{i}} 
= \vC{\nu}{i} \textrm{ if } \svA + \elementaryVectorNoSize{i} \in \stateSpace 
$,
and to states $\svA - \elementaryVectorNoSize{i}$, $i = 1, \ldots, N$, with rate
$
\matrixElement{Q}{\svA}{\svA - \elementaryVectorNoSize{i}} 
= \vC{\mu}{i} \textrm{ if } \svA - \elementaryVectorNoSize{i} \in \stateSpace 
$.
All other off-diagonal elements of $Q$ are set to zero, which (for the Rydberg gas model) means that we neglect multiphoton processes. For completeness, we note that the diagonal elements are given by $\matrixElement{Q}{\svA}{\svA} = - \sum_{\svB \neq \svA} \matrixElement{Q}{\svA}{\svB}$.
We conclude that the stochastic process described by the generator matrix $Q$, which we denote by $\process{ \vect{X}(t) }{ t \geq 0 }$, is a model for wireless random-access networks, as well as Rydberg gases.


We now investigate steady states of the Rydberg gas, using our understanding of wireless networks. The equilibrium fraction of time that the system spends in state $\svA$ is given by
\begin{equation}
\vC{\pi}{\svA}( \vect{\nu}, \vect{\mu} ) 
= \frac{1}{ \normalizationConstant( \vect{\nu}, \vect{\mu} ) } \prod_{i=1}^N \Bigl( \frac{ \vC{\nu}{i} }{ \vC{\mu}{i} } \Bigr)^{ \vC{\sA}{i} },
\quad \svA \in \stateSpace,
\label{eqn:Equilibrium_distribution}
\end{equation}
with $\normalizationConstant( \vect{\nu}, \vect{\mu} )$ denoting the normalization constant. The equilibrium distribution depends solely on the ratios $\vC{\nu}{i} / \vC{\mu}{i}$, and proving that it is in fact the equilibrium distribution can be done by observing that it satisfies the detailed balance equations \cite{kelly_reversibility_2011}, $\vC{\pi}{\svA} \matrixElement{Q}{\svA}{\svB} = \vC{\pi}{\svB} \matrixElement{Q}{\svB}{\svA}$, for all $\svA, \svB \in \stateSpace$.

Consider the special case in which all particles make their transition at the same rate, and set $\vC{\nu}{i} = \nu$ and $\vC{\mu}{i} = \mu$ for $i = 1, \ldots, N$ accordingly. When $\nu / \mu \rightarrow \infty$, the equilibrium probability of observing the system in state $\svA \in \stateSpace$ converges to
\begin{equation}
\vC{\pi}{\svA}( \vect{\nu}, \vect{\mu} ) 
= 
\frac{1}{ \normalizationConstant( \vect{\nu}, \vect{\mu} ) } \Bigl( \frac{\nu}{\mu} \Bigr)^{ \sum_{i=1}^N \vC{\sA}{i} }
\rightarrow
\frac{ \indicator{ \svA \in \mathcal{I} } }{ \cardinality{ \mathcal{I} } },
\label{eqn:Equilibrium_distribution_when_rho_is_large}
\end{equation}
where $\mathcal{I}$ denotes the collection of maximum independent sets of $\stateSpace$. In the present context, a maximum independent set is a configuration in which the largest number of particles are active, i.e.\ in the Rydberg state. 
We call these configurations dominant, because the probability of observing a dominant configuration $\svA \in \mathcal{I}$, $\pi_{\svA}$, is large compared to (i) the probability of observing a configuration $\svB \notin \mathcal{I}$, $\pi_{\svB}$, when $\nu \gg \mu$, and (ii) the probability of observing any but the dominant configuration, $\sum_{\svB \notin \mathcal{I}} \pi_{\svB}$, when $\nu \gg N \mu$.

\begin{figure}[hbt]
\begin{center}
\subfigure{
\includegraphics[height=0.206\columnwidth]{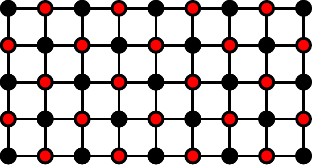}
}
\subfigure{
\includegraphics[height=0.206\columnwidth]{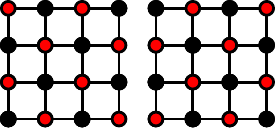}
}
\end{center}
\vspace{-0.75cm}
\caption{Dominant configurations in a (left) $9 \times 5$ lattice and (middle, right) $4 \times 4$ lattice with nearest-neighbor blocking.}
\label{fig:Maximal_independent_set_of_an_9x5_lattice}
\vspace{-1em}
\end{figure}

The active particles in dominant configurations typically form patterns, which resemble crystal structures. Consider for instance an $n \times m$ lattice of particles exhibiting nearest-neighbor blocking, where $n, m \in \naturalNumbersPlus$. For such networks, the active particles in the dominant configuration follow a checkerboard pattern, as illustrated in \refFigure{fig:Maximal_independent_set_of_an_9x5_lattice}. When both $n$ and $m$ are even, two dominant configurations exist, which we henceforth refer to as the even and odd configuration.

Our analysis reveals that when $\RabiFrequency_{\textrm{r}} \ll \RabiFrequency_{\textrm{e}}$, the Rydberg gas spends more time in a dominant configuration than in another configuration. 
The time it takes for the system to switch between different dominant configurations is related to the mixing time of the system, i.e. the time required for the Markov process to get sufficiently close to stationarity \cite{levin_markov_2009}. Depending on the topology, the mixing time can be large when $\RabiFrequency_{\textrm{r}} \ll \RabiFrequency_{\textrm{e}}$, implying that once the system is in a dominant configuration, it tends to stay there for a long time.
It is noteworthy that simulations of a driven dissipative Rydberg gas confirmed the formation of crystalline structures in \cite{honing_steady-state_2013,Hu13,Petrosyan13}, and here we have explained how such formations appear using our connection to wireless networks.


We are also able to investigate the time $\tau$ it takes until the process reaches a dominant configuration. The hitting time $\tau$ of the dominant configuration is the first moment at which the system reaches the even or odd dominant configuration. The random variable $\tau$ is of interest, because it is a measure for how long the experimentalist has to wait before a dominant configuration has appeared. 

\begin{figure}[!hbtp]
\begin{center}
\includegraphics[width=0.995\columnwidth]{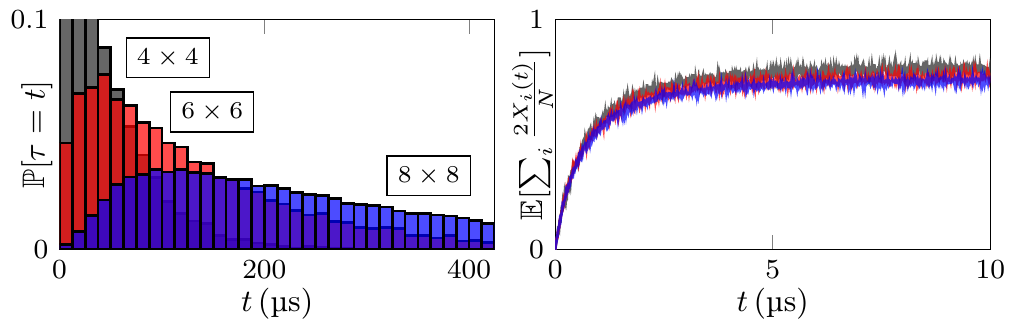}
\vspace{-0.75cm}
\caption{\textrm{Histograms of the hitting time distributions (left), and normalized average number of excited particles (right) for lattices of sizes $4 \times 4$, $6 \times 6$, and $8 \times 8$. Here, $\decayRate = 2 \pi \cdot \SI{6}{\mega\hertz}$, $\RabiFrequency_{\textrm{e},i} = 2 \pi \cdot \SI{3}{\mega\hertz}$ and $\RabiFrequency_{\textrm{r},i} = 2 \pi \cdot \SI{1}{\mega\hertz}$.}}
\label{fig:Histogram_of_hitting_time__Even}
\end{center}
\vspace{-1em}
\end{figure}

To illustrate this, we have simulated sample paths of $\vect{X}(t)$ on even $n \times n$ lattice topologies. Histograms of the hitting time distributions for grids of several sizes are shown in \refFigure{fig:Histogram_of_hitting_time__Even}, as well as the normalized average number of excited particles. Note that the average hitting time increases as lattices become larger.

We now describe a wireless network algorithm in the context of Rydberg gases, to determine Rabi frequencies (laser intensities) such that particles in the Rydberg gas are excited with specified target excitation probabilities. The algorithm was developed in \cite{jiang_distributed_2008} to achieve maximum throughput in wireless networks in a distributed fashion, and was later generalized for implementation in product-form networks \cite{sanders_online_2012}. 
In Supplementary material \cite{supplementary_material}, we provide a short discussion of the algorithm in its original context, and we explain that the algorithm is solving an inversion problem that can be NP hard.

The wireless network algorithm can be applied to the Rydberg atoms by iteratively setting
\begin{equation}
| \itr{ \RabiFrequency_{\textrm{e},i} }{n+1} | = | \itr{ \RabiFrequency_{\textrm{e},i} }{n} | \exp \Bigl( - \frac{1}{2} \itr{a}{n+1} ( \itr{\vC{\hat{\theta}}{i}}{n+1} - \vC{\phi}{i} ) \Bigr), \label{eqn:Rydberg_gas_algorithm}
\end{equation}
for atoms $i = 1, \ldots, N$. Here, $n \in \naturalNumbersPlus$ indexes each iteration, and the $\itr{a}{n}$ denote algorithm step sizes that are typically chosen as a decreasing sequence. The $\itr{ \vC{\hat{\theta}}{i} }{n+1}$ denote empirically obtained estimates of the probabilities of observing atom $i = 1, \ldots, N$ in the Rydberg state, $\vC{\theta}{i}$, and $\vC{\phi}{i}$ denotes the target probability of observing atom $i$ in the Rydberg state. 
The algorithm in \refEquation{eqn:Rydberg_gas_algorithm} seeks $\criticalpoint{ \vect{\RabiFrequency}_{\textrm{e}} }$ such that $\vect{\theta}( \criticalpoint{ \vect{\RabiFrequency}_{\textrm{e}} }, \vect{ \RabiFrequency_{\textrm{r}} } ) = \vect{\phi}$.

In wireless networks, an estimate $\vC{\hat{\theta}}{i}$ can be obtained through online observation of a transmitter's activity [\emph{Supplemental material}, Eq.~(1)]. Experimentally observing the evolution of a particle system through time however is difficult.
Instead, we can (i) determine an estimate $\vect{\hat{\theta}}$ of $\vect{\theta}$ using simulation, or (ii) use repeated experimentation to determine an estimate $\vect{\hat{\theta}}$ of $\vect{\theta}$. With the latter approach, we forego our mathematical guarantee of convergence, but the design principles that guaranteed the convergence in the former method still hold. That is, we need to improve the quality of $\vect{\hat{\theta}}$ as the number of iterations $n$ increases.

For every $n$th iteration of the algorithm, we can for example reinitialize the process $\itr{m}{n}$ times and determine the state the process is in at some time $\itr{T}{n}$. Denoting these samples by $\itr{\vC{X}{i}}{n,s} ( \itr{T}{n} )$, with $n \in \naturalNumbersPlus$ and $s \in \{ 1, \ldots, \itr{m}{n} \}$, we can calculate
\begin{equation}
\itr{\vC{\hat{\theta}}{i}}{n} 
= \frac{1}{ \itr{m}{n} } \sum_{s=1}^{ \itr{m}{n} } \indicator{ \itr{\vC{X}{i}}{n,s} ( \itr{T}{n} ) = 1 },
\,\,\, i = 1, \ldots, N,
\label{eqn:Emperical_estimate_of_the_probability_that_particle_i_is_excited_with_repeated_experimentation}
\end{equation}
which, for sufficiently large $\itr{T}{n}$ and $\itr{m}{n}$, provides an estimate of the equilibrium probability that particle $i$ is in the Rydberg state. Intuitively, we expect that $\itr{T}{n}$ should be at least of the order of the mixing time (that is to say, the system should be close to equilibrium).

\begin{figure}[!hbtp]
\begin{center}
\includegraphics[width=0.99\columnwidth]{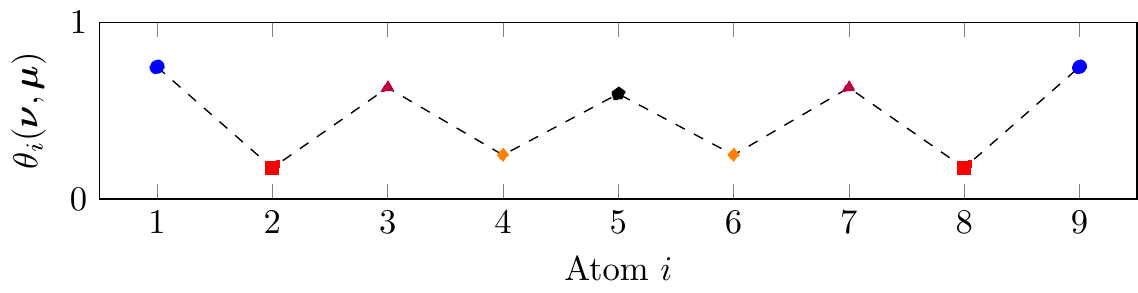}
\vspace{-0.75cm}
\caption{\textrm{The $\vC{\theta}{i}( \vect{\nu}, \vect{\mu} )$ for $N=9$, $b=1$, and $\vC{\nu}{i} / \vC{\mu}{i} = 10$.}}
\label{fig:Throughput_of_nine_atoms_on_a_line_with_nu_10_and_mu_1}
\end{center}
\vspace{-1em}
\end{figure}

As an example, we focus on a system of $i = 1, \ldots, N$ atoms positioned on a line, that block the first $b$ neighbors on both sides. We consider the problem of determining $\vect{ \RabiFrequency_{\textrm{e}} }$ such that each atom is excited with equal probability $\phi \in (0,1)$. This problem is nontrivial because the atoms at the border have fewer neighbors that block them and are therefore excited with higher probability. Moreover, this effect propagates through the system, which can be verified by an analytical evaluation of the probabilities of observing atom $i$ in the Rydberg state,
\begin{equation}
\vC{\theta}{i}(\vect{\nu}, \vect{\mu}) = \sum_{ \svA \in \stateSpace } \vC{\sA}{i} \vC{\pi}{\svA}(\vect{\nu}, \vect{\mu}),
\quad i = 1, \ldots, N,
\label{eqn:Throughput}
\end{equation}
as shown in \refFigure{fig:Throughput_of_nine_atoms_on_a_line_with_nu_10_and_mu_1}. 

We consider this particular example because we can again utilize our connection to wireless networks and provide an analytical expression for $\criticalpoint{ \vect{ \RabiFrequency }_{\textrm{e}} }$. As shown in \cite{van_de_ven_spatial_2012}, we need to set
\begin{equation}
\Bigl( \frac{ \criticalpoint{ \RabiFrequency_{\textrm{e},i} } }{ \RabiFrequency_{\textrm{r},i} } \Bigr)^2 = \frac{ \phi }{ 1 - (1 + b) \phi } \Bigl( \frac{ 1 - b \phi }{ 1 - (1 + b) \phi } \Bigr)^{ w(i) - w(1) } 
\label{eqn:VanDeVenSolution}
\end{equation}
for $i = 1, \ldots, N$, in order to have $\vectComponent{\theta}{i}( \criticalpoint{ \vect{ \RabiFrequency }_{\textrm{e}} }, \vect{ \RabiFrequency_{\textrm{r}} } ) = \phi$ for $i = 1, \ldots, N$. Here, $w(i) = \min \{ i + b, N \} - \max \{ 1, i - b \}$ denotes the number of other atoms that atom $i$ blocks if it is excited.

\begin{figure}[!hbtp]
\begin{center}
\includegraphics[width=0.995\columnwidth]{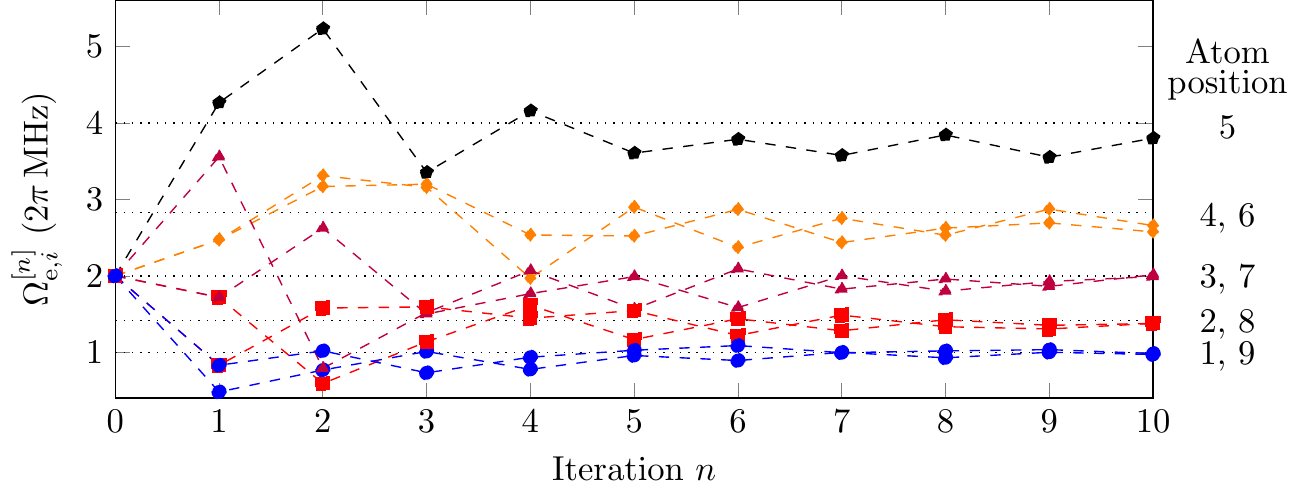}
\vspace{-0.75cm}
\caption{\textrm{
Algorithm output when $N = 9$, $b = 4$, $\decayRate = 2\pi \cdot \SI{6}{\mega\hertz}$, and $\RabiFrequency_{\textrm{r},i} = 2\pi \cdot \SI{1}{\mega\hertz}$. The dotted lines indicate $\criticalpoint{ \vect{\RabiFrequency}_{\textrm{e}} }$.}}
\label{fig:Rydberg_gas_simulation}
\end{center}
\vspace{-1em}
\end{figure}

In order to illustrate the algorithm applied to this system, we utilize the following simulation procedure. We repeatedly simulate the Rydberg gas by generating sample paths $\itr{\vect{X}}{n,s}(t)$ using the generator matrix in \refEquation{eqn:Rate_equations}. Subsequently, we calculate an estimate of the excitation probabilities through \refEquation{eqn:Emperical_estimate_of_the_probability_that_particle_i_is_excited_with_repeated_experimentation}, and update the Rabi frequencies according to the algorithm in \refEquation{eqn:Rydberg_gas_algorithm}. In every $n$th iteration of our algorithm, we set the maximum simulation time to $\itr{T}{n} = \SI{250}{\micro\second}$, produce $\itr{m}{n}= 25n^2$ samples, and choose step size $\itr{a}{n} = 100/(10 + \sqrt{n})$. The target excitation probability of the algorithm is set to $\phi = 1/6$. 
The resulting Rabi frequencies are shown in \refFigure{fig:Rydberg_gas_simulation}, and approach the exact solution given by \refEquation{eqn:VanDeVenSolution},
$\criticalpoint{ \vect{\RabiFrequency}_{\textrm{e}} } = ( 1$, $\sqrt{2}$, $2$, $2 \sqrt{2}$, $4$, $2 \sqrt{2}$, $2$, $\sqrt{2}$, $1 \transpose{)} \cdot \SI{2\pi}{\mega\hertz}$.

\begin{figure}[!hbtp]
\begin{center}
\includegraphics[width=0.995\columnwidth]{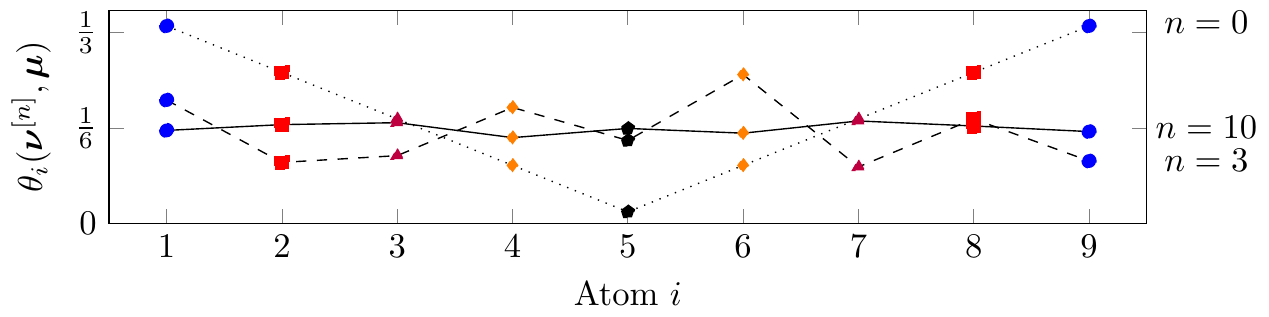}
\vspace{-0.75cm}
\caption{\textrm{The excitation probabilities $\theta_i(\iterand{\vect{\nu}}{n},\vect{\mu})$ after iterations $n = 0$, $3$, and $10$.}}
\label{fig:Throughput_of_nine_atoms_on_a_line}
\end{center}
\vspace{-1em}
\end{figure}

The excitation probabilities approach the target $\phi$, which can be verified by evaluating \refEquation{eqn:Throughput} after several iterations, as shown in \refFigure{fig:Throughput_of_nine_atoms_on_a_line}.

By manipulating excitation probabilities, we control the populations of mixed states. This can be of interest to (for example) mixed state quantum computing, which lies in between classical computing and quantum computing based on pure, entangled states \cite{Siomau11,Blume02}. Creating mixed states can also be a first step towards efficient preparation of large qubit entangled states.


In conclusion, we studied the relations between a physical model of ultracold Rydberg atoms and a stochastic process that models certain wireless random-access networks. This allowed us to identify interesting connections between research fields in physics and mathematics, and to transfer techniques and insights 
to the realm of Rydberg gases. 
Our approach can be applied to many other particle systems and stochastic processes as well. Furthermore, the algorithm can be applied to a much larger class of product-form networks, with different adjustable parameters \cite{sanders_achievable_2012,sanders_online_2012}. Whenever dynamical systems are well described using rate equations, it can be worthwhile to explore possible relations with stochastic processes and cross-pollinate ideas.

\appendix

\begin{acknowledgments}
This research was financially supported by an ERC Starting Grant, as well as The Netherlands Organization for Scientific Research (NWO), and is part of the research program of the Foundation for Fundamental Research on Matter (FOM). 
The authors are grateful for the support from Sem Borst and Johan van Leeuwaarden.
\end{acknowledgments}

\bibliographystyle{apsrev}
\bibliography{Bibliography}

\section{Supplementary material}

\input{Supplementary_material__Include_file}

\end{document}

%% file: Supplementary_material__Include_file.tex
We propose using a wireless network algorithm to achieve target excitation probabilities in a Rydberg gas. The algorithm will calculate Rabi frequencies so that the probability that particle $i$ is in the Rydberg state is equal to a target probability $\vectComponent{\phi}{i} \in (0,1)$ that we can specify for every particle $i = 1, \ldots, N$. In formula, the probability that particle $i$ is in the Rydberg state is given by [\emph{Wireless network control of interacting Rydberg atoms}, Eq.~(9)]. The dependence on the system parameters $\vect{\nu}$, $\vect{\mu}$ is emphasized by writing the equilibrium distribution as a function of $\vect{\nu}$ and $\vect{\mu}$.

In the context of a wireless random-access network, this performance measure is called normalized throughput and may in fact not be fair for end-users. We have for instance seen that when $\nu \gg \mu$ (a heavily loaded system), certain transmitters are active for much larger fractions of time than other transmitters, recall [\emph{Wireless network control of interacting Rydberg atoms}, Eq.~(6)]. This leads to starvation effects and unfairness in wireless random-access networks, which has been studied in for example \cite{wang_throughput_2005,durvy_modeling_2007,durvy_self-organization_2009,denteneer_IEEE_2008}. In wireless networks, it is therefore desirable to find system parameters $\vect{\criticalpoint{\nu}}$ such that $\vect{\theta}(\vect{\criticalpoint{\nu}}, \vect{\mu}) = \vect{\phi}$, where $\vect{\phi}$ denotes fairer probabilities that transmitters are active. This problem is difficult to solve analytically, as well as numerically. For example, in order to evaluate [\emph{Wireless network control of interacting Rydberg atoms}, Eq.~(6)], we need to determine all maximum independent sets of a graph, which is a well-known NP-hard problem.

Recent studies of the wireless network model have led to the development of distributed algorithms to solve $\vect{\theta}(\vect{\criticalpoint{\nu}}, \vect{\mu}) = \vect{\phi}$ for $\vect{\criticalpoint{\nu}}$, without needing to numerically evaluate the equilibrium distribution. Let $0 = \itr{t}{0} < \itr{t}{1} < \ldots$ denote points in time, constituting time slots $[ \itr{t}{n}, \itr{t}{n+1} ]$. At time $\itr{t}{n+1}$, the end of the $(n+1)$-th time slot, each transmitter $i = 1, \ldots, N$ calculates the empirical estimate
\begin{equation}
\itr{\vC{\hat{\theta}}{i}}{n+1} 
= \frac{1}{ \itr{t}{n+1} - \itr{t}{n} } \int_{\itr{t}{n}}^{\itr{t}{n+1}} \indicator{ \itr{\vC{X}{i}}{n}(t) = 1 } dt \label{eqn:Emperical_estimate_of_the_probability_that_particle_i_is_excited}
\end{equation}
of the probability that it was active. Here, $\itr{\vect{X}}{n}(t) : [ \itr{t}{n}, \itr{t}{n+1} ) \rightarrow \stateSpace$ denotes a realized sample path of the stochastic process that describes the wireless network operating with rates $\vect{\mu}$ and $\itr{\vect{\nu}}{n}$ during time interval $[ \itr{t}{n}, \itr{t}{n+1} )$. Keeping $\vect{\mu}$ fixed, transmitters $i = 1, \ldots, N$ next update
\begin{equation}
\itr{\vC{\nu}{i}}{n+1} = \itr{\vC{\nu}{i}}{n} \exp \bigl( - \itr{a}{n+1} ( \itr{\vC{\hat{\theta}}{i}}{n+1} - \vC{\phi}{i} ) \bigr),
\quad n \in \naturalNumbersZero,
\label{eqn:Wireless_network_algorithm}
\end{equation}
where we denote the step size of the algorithm by $\itstep{n}$. This update procedure is then repeated. It is noteworthy that the right-hand side in \refEquation{eqn:Wireless_network_algorithm} is independent of $\vC{\hat{\theta}}{j}$ and $\vC{\phi}{j}$ for $j \neq i$, allowing each transmitter to base its decisions on local information only.

If the step sizes $\itr{a}{n}$ and time-slots are chosen appropriately, $\itr{\vect{\nu}}{n}$ converges to $\criticalpoint{\vect{\nu}}$ with probability one \cite{jiang_distributed_2008,sanders_online_2012}. The idea is to increase $\itr{t}{n+1} - \itr{t}{n}$ as $n$ increases such that better estimates of $\itr{\vect{\theta}}{n}$ are obtained, while slowly decreasing $\itr{a}{n}$ to prevent poor decision-taking, but not too slow and be unresponsive. The precise form of the conditions depends on the specific modelling conditions. If the set of feasible configurations is finite and if one projects the outcome of the algorithm to a compact set, which are reasonable assumptions in practice due to typically having finite capacities and resources, it suffices to choose $\itr{a}{n}$ and $\itr{t}{n}$ such that $\sum_{n = 1}^\infty \itstep{n} = \infty$, $\sum_{n = 1}^\infty (\itstep{n})^2 < \infty$ and $\sum_{n = 1}^\infty \itstep{n} / ( \itr{t}{n} - \itr{t}{n-1} ) < \infty$, e.g.~$\itstep{n} = n^{-1}$, $\itr{t}{n} - \itr{t}{n-1} = n$.


A technical difficulty is determining whether there even exists a finite $\criticalpoint{\vect{\nu}}$ such that $\vect{\theta}(\criticalpoint{\vect{\nu}}, \vect{\mu}) = \vect{\phi}$. If such $\criticalpoint{\vect{\nu}}$ exists, we call $\vect{\phi}$ achievable. We readily obtain the answer by again looking at known results for the wireless network model \cite{jiang_distributed_2008}, which have also later been generalized for product-form networks \cite{sanders_achievable_2012}. In short, any
\begin{align}
\vect{\phi} 
\in 
\bigl\{ \sum_{\svA \in \stateSpace} \vC{\alpha}{\svA} \svA \big| \vect{\alpha} \in (0,1)^{ \cardinality{\stateSpace} }, \transpose{ \vect{\alpha} } \vectones{ \cardinality{\stateSpace} } = 1 \bigr\}
\label{eqn:Achievable_region}
\end{align}
is achievable. Here, $\vectones{ \cardinality{\stateSpace} }$ denotes the $\cardinality{\stateSpace}$-dimensional vector that contains all ones. \refEquation{eqn:Achievable_region} is a convex hull of the configurations in $\stateSpace$.